\newcommand{\be}[0]{\begin{equation}}
\newcommand{\ee}[0]{\end{equation}}
\newcommand{\ba}[0]{\begin{eqnarray}}
\newcommand{\ea}[0]{\end{eqnarray}}

\newcommand\GeV{\,\mbox{GeV}}
\newcommand\MeV{\,\mbox{MeV}}

\def\a{\alpha }

\documentclass[report]{JHEP3}

\usepackage{epsfig,multicol}

\newcommand\fverb{\setbox\pippobox=\hbox\bgroup\verb}
\newcommand\fverbdo{\egroup\medskip\noindent%
            \fbox{\unhbox\pippobox}\ }
\newcommand\fverbit{\egroup\item[\fbox{\unhbox\pippobox}]}
\newbox\pippobox
\title{The Jacobi Polynomials QCD analysis for the polarized structure function
}

\author{S. Atashbar Tehrani $^{(a,c)}$ and Ali N. Khorramian $^{(b,c)}$

\\
\small{\it
$^{(a)}$ Physics Department, Persian Gulf University, Boushehr, Iran\\
$^{(b)}$ Physics Department, Semnan University, Semnan, Iran \\
$^{(c)}$ Institute for Studies in Theoretical Physics and
Mathematics (IPM), \\ \hspace{.48 cm}P.O.Box 19395-5531, Tehran, Iran \\
 }
\\

    E-mail:   \email{atashbar@ipm.ir}, \email{khorramiana@theory.ipm.ac.ir}}
\received{}        %%
\revised{}
\accepted{}        %% These are for published papers.
\abstract{We present the results of our QCD analysis for polarized
quark distribution and structure function $xg_1 (x,Q^2)$. We use
very recently experimental data to parameterize our model. New
parameterizations are derived for the quark and gluon
distributions for the kinematic range $x\;\epsilon\;[10^{-8},1]$,
$Q^2\;\epsilon\;[1,10^6]$ GeV$^2$. The analysis is based on the
Jacobi polynomials expansion of the polarized structure functions.
Our calculations for polarized parton distribution functions based
on the Jacobi polynomials method are in good agreement with the
other theoretical models. The values of $\Lambda_{QCD}$ and
$\alpha_s(M_z)$ are determined.}

\begin{document}
\section{Introduction}
The nature of the short-distance structure of polarized nucleons is
one of the central questions of present day hadron physics. For more
than sixteen years, polarized inclusive deep inelastic scattering
has been the main source of information on how the individual
partons in the nucleon are polarized at very short distances.

The theoretical and experimental status on the spin structure of
the nucleon  has been discussed in great detail in several recent
reviews (see, e.g.,
Refs.~\cite{Anselmino:1994gn,Lampe:1998eu,Hughes:1999wr,filippone-01}.
During the recent years several comprehensive analysis of the
polarized deep inelastic scattering (DIS) data, based on
next-to-leading-order quantum chromodynamics, have appeared in
Refs.[5-29]. In these analysis the polarized parton density
functions (PPDFs) are either written in terms of the well-known
parameterizations of the unpolarized PDFs or parameterized
independently, and the unknown parameters are determined by
fitting the polarized DIS data.

Determination of  parton distributions in a nucleon in the
framework of quantum chromodynamics (QCD)  always involves some
model-dependent procedure. Instead of relying on mathematical
simplicity as a guide, we take a viewpoint in which  the physical
picture of the nucleon structure is emphasized. That is, we
consider the model for the nucleon which is compatible with the
description of the bound state problem in terms of three
constituent quarks. We adopt the view that these
 constituent quarks in the scattering problems should be regarded as the valence quark
clusters rather than point-like objects. They have been referred
to as $\it{valons}$. In the valon model, the proton consists of
two ``up'' and one ``down'' valons. These valons thus, carry the
quantum numbers of the respective valence quarks. Hwa and et al.
[30-38] found evidence for the valons in the deep inelastic
neutrino scattering data, suggested their existence and applied it
to a variety of phenomena. Hwa \cite{Hwa:1979pn} has also
successfully formulated a treatment of the low-$p_{T}$ reactions
based on a structural analysis of the valons.  In
\cite{Hwa:2002mv}
 Hwa and Yang refined the idea of the valon model and
extracted new results for the valon distributions. In
\cite{Arash:2003ci,Arash:2003js,Arash:2003eh} unpolarized PDFs and
hadronic structure functions in the NLO approximation were
extracted.

In Ref.~\cite{Khorramian:2004ih} the polarized valon model is
applied to determine the quark helicity distributions and
polarized proton structure functions in the NLO approximation by
using the Bernstein polynomial approach.

The extraction of the quark helicity distributions is one of the
main tasks of the semi-inclusive deep inelastic scattering (SIDIS)
experiments (HERMES \cite{Airapetian:2004zf}, COMPASS
\cite{compass}, SMC\cite{Adeva:1997qz}) with the polarized beam
and target. Recently in Ref. \cite{Mirjalili:2006hf} the polarized
valon model was applied and analyzed the flavor-broken light sea
quark helicity distributions with the help of a Pauli-blocking
ansatz. The results reported of this paper is good agreement with
the HERMES experimental data for the quark helicity distributions
in the nucleon for up, down, and strange quarks from
semi-inclusive deep-inelastic scattering \cite{Airapetian:2004zf}.
The reported results of Ref.~\cite{Airapetian:2004zf} is based on the Bernstein polynomial
expansion method in the polarized valon model framework.

Since very recently experimental data are available from the HERMES
collaboration \cite{unknown:2006vy} for the spin structure function
$g_1$, therefore there is enough motivation to study and utilize the
spin structure and quark helicity distributions extracted
via the phenomenological model. Since these recently experimental data
are just for different value of $Q^2$ and not fixed $Q^2$, one can not use the
the Bernstein polynomial expansion method. Because in this method we need to use
experimental data for each bin of $Q^2$ separately. So it seems suitable, to use the Jacobi
polynomial expansion method.\\

In this paper we use the idea of the polarized valon model to obtain
the PPDFs in the LO and NLO approximations. The results of the present analysis is based on
the Jacobi polynomials expansion of the polarized structure function.\\

The plan of the paper is to give an introduction to the polarized
valon distributions in Section 2. Parametrization of parton
densities are written down in this section. In Section 3 we
present a brief review of the theoretical background of the QCD
analysis in two loops. The method of the QCD analysis of polarized
structure function, based on Jacobi polynomials are written down
in Section 4. A description of the procedure of the QCD fit of
$g_1$ data are illustrated in Section~5. Section 6 contains a
results and conclusions of the QCD analysis.
\\
%\section{Formalism}

\section{Polarized valon distributions }
The idea of nucleon as a bound state of three quarks was presented
for the first time in Ref. \cite{Altarelli:1973ff}. On the other
hand the similar idea, which is called valon model was presented
in Ref. \cite{Hwa:1980mv}. According to the valon model framework,
one can assume a simple form for the exclusive valon distribution
which facilitate the phenomenological analysis as follows
\begin{equation}
G_{UUD/p}(y_{1},y_{2},y_{3})=g (y_{1}y_{2})^{a }y_{3}^{b }\delta
(y_{1}+y_{2}+y_{3}-1)\;,
\end{equation}
where $y_{i}$ is the momentum fraction of the $i$'th valon. The
$U$ and $D$ type inclusive valon distributions can be  obtained by
double integration over the unspecified variables
\begin{eqnarray}
G_{U/p}(y) &=&\int dy_{2}\int dy_{3}G_{UUD/p}(y,y_{2},y_{3}) \\
&=&g B(a +1,b +1)y^{a }(1-y)^{a +b +1}\;, \nonumber
%\\
\end{eqnarray}
\begin{eqnarray}
G_{D/p}(y) &=&\int dy_{1}\int dy_{2}G_{UUD/p}(y_{1},y_{2},y) \\
&=&g B(a +1,a +1)y^{b }(1-y)^{2a +1}\;. \nonumber
\end{eqnarray}
The normalization parameter $g$ has been fixed by \be
\int^{1}_{0}G_{U/p}(y)dy=\int^{1}_{0}G_{D/p}(y)dy=1\;, \ee \\
and is equal to $ g=[B(a +1,b +1)B(a +1,a +b +2)]^{-1}$, where
$B(m,n)$ is the Euler-Beta function. R.C. Hwa and C. B. Yang
\cite{Hwa:2002mv} have recalculated the unpolarized valon
distribution in the proton with a new set of parameters. The new
values of $a $ , $b $ are found to be $a=1.76 $ and
$b =1.05 $.\\

To describe the quark distribution $q(x)$ in the valon model, one
can try to relate the polarized quark distribution functions
$q^\uparrow$ or $q^\downarrow$ to the corresponding valon
distributions $G^\uparrow$ and $G^\downarrow$. The polarized valon
can still have the valence and sea quarks that are polarized in
various directions, so long as the net polarization is that of the
valon. When we have only one distribution $q(x,Q^2)$ to analyze,
it is sensible to use the convolution in the valon model to
describe the proton structure in terms of the valons. In the case
that we have two quantities, unpolarized and polarized
distributions, there is a choice of which linear combination
exhibits more physical contents. Therefore, in our calculations we
assume a linear combination of $G^\uparrow$ and $G^\downarrow$ to
determine respectively the unpolarized ($G$) and polarized
($\Delta G$) valon distributions .

According to unpolarized and polarized valon model framework we
have \cite{Khorramian:2004ih}
 \ba q_{i/p}(x,Q^{2})= \sum_{j}\int_{x}^{1}
q_{i/j}(\frac{x}{y},Q^{2})G_{j/p}(y)\frac{dy}{y}\;,
\label{Convunp} \ea
and
 \ba \Delta
q_{i/p}(x,Q^{2})= \sum_{j}\int_{x}^{1}  \Delta
q_{i/j}(\frac{x}{y},Q^{2})\Delta G_{j/p}(y)\frac{dy}{y}\;.
\label{Convp} \ea where $ q_{i}\equiv
q^\uparrow_{i}+q^\downarrow_{i}$ and $ \Delta q_{i}\equiv
q^\uparrow_{i}-q^\downarrow_{i}$
 and also $G_{j}\equiv\;G^\uparrow_{j} +\;G^\downarrow_{j}$ and
 $\Delta G_{j}\equiv\;G^\uparrow_{j}
-\;G^\downarrow_{j}$. As we can  similarly see for the unpolarized
case, the polarized quark distribution can be related to a
polarized valon distribution in Eq.~(\ref{Convp}).

The polarized parton distribution of the polarized constituent quarks in the Eq.~(\ref{Convp})
is dependent on a probe \cite{Khorramian:2004ih}. It is assumed in more detail that the
polarized constituent quarks distribution, that is, $\Delta q_{i/j}(z=\frac{x}{y},Q^{2})$
in the scale of $Q_0^2$ is equal to $\Delta q_{i/j}(z,Q_0^{2})=\delta (1-z)$. From Eq.~(\ref{Convp})
 now is obvious that
the polarized quark distribution in the scale of $Q_0^2$ is equal to
polarized valon distribution.

Here we want to use the polarized valon distributions which
introduced in \cite{Khorramian:2004ih}. According to improve
polarized valon picture the nonsinglet polarized valon
distribution functions are as following

\begin{equation}
\Delta G^{}_{j}(y)=\Delta {\cal {W}}^{}_{j}(y)\times G_{j}(y)\;,
\label{delGNS}
\end{equation}
where \begin{equation} \label{waight1} \Delta {\cal
{W}}^{}_{j}(y)=\xi_j A_{j}  y^{\alpha _{j}}(1-y)^{\beta
_{j}}(1+\gamma _{j}y+\eta _{j}y^{0.5})\;,
\end{equation}
the subscript $j$ refers to $U$ and $D$-valons. The motivation for
choosing this functional form is that the $y^{\alpha _{j}}$ term
controls the low-$y$ behavior valon densities, and $(1-y)^{\beta
_{j}}$ terms the large-$y$ values. The remaining polynomial factor
accounts for the additional medium-$y$ values. The normalization
constants $A_j$  for $U$ and $D$-valons are as following
\begin{eqnarray}
{A_{_U}}^{-1}&=&\left[B(a+{\alpha_{_U}},2+a+b+{\beta_{_U}})+{\eta
_{_U}}B(0.5+a+{\alpha_{_U}},2+a+b+{\beta_{_U}})\nonumber \right.\\
&&\left.+{\gamma_{_U}}B(1+a+{\alpha
_{_U}},2+a+b+{\beta_{_U}})\right ]/{B(1+a,2+a+b)}\;,
 \end{eqnarray}and
\begin{eqnarray}
{A_{_D}}^{-1}&=&\left[B(b+\alpha_{_D},2+2
a+{\beta_{_D}})+\eta_{_D} B(0.5+b+{\alpha_{_D}},{2}+{2}
a+{\beta_{_D}})\nonumber \right.\\&&\left.+\gamma_{_D}
B(1+b+\alpha_{_D},2+2 a+\beta_{_D})\right ]/B(b+1,2a+2)\;.
\end{eqnarray}
These quantities are chosen such that the $\xi_j$  are the first moments of $\Delta
G_{j}(y)$, $\xi_j=\int_0^1 dy \Delta G_{j}(y)$. Here $B(a,b)$ is
the Euler Beta--function being related to the $\Gamma$--function.

In the present approach the QCD--evolution equations are solved in
{\sc Mellin}--$N$ space as will describe in section 3. The {\sc
Mellin}--transform of the valon densities is performed and {\sc Mellin}--$N$
 moments are calculated for complex arguments $N$ by
$\int_0^1y^{N-1}\Delta G_{j/p}(y)dy$.\\

As seen from Eq.~(\ref{delGNS}) there are five parameters for each
valon distribution. To meet both the quality of the present data
and the reliability of the fitting program, the number of
parameters has to be reduced. Assuming $SU(3)$ flavor symmetry and
a flavor symmetric sea one only has to derive one general
polarized sea--quark distribution. The first moments of the
polarized valon distributions $\Delta G_{_U}$ and $\Delta G_{_D}$
can be fixed by the $SU(3)$ parameters $F$ and $D$ as measured in
neutron and hyperon $\beta$--decays according to the relations~:

\vspace*{-0.75cm}
\begin{eqnarray}
2\xi_{_U} - \xi_{_D} & = & F + D~,      \nonumber     \\
2\xi_{_U} + \xi_{_D} & = & 3F - D~. \label{xi}
\end{eqnarray}

\noindent The factor 2 in Eq.~(\ref{xi}) is due  to the existence
of two-$U$ type valons. A re-evaluation of $F$ and $D$ was
performed in Ref.~\cite{Goto:1999by} on the basis of updated
$\beta$--decay constants \cite{Caso:1998tx} leading to
\begin{eqnarray}
2\xi_{_U} &=& ~~0.926 \pm 0.014~, \nonumber \\
\xi_{_D} &=& -0.341 \pm 0.018~.
\end{eqnarray}
This choice reduces the number of parameters to be fitted for each
nonsinglet  valon density to four.\\

On the other hand the singlet polarized valon distribution which
is defined in \cite{Khorramian:2004ih} is as following

\begin{equation}
\Delta G^{'}_{j}(y)=\Delta {\cal {W}}^{'}_{j}(y)\times G_{j}(y)
\label{delGS}
\end{equation}
 where $\Delta {\cal
{W}}^{'}_{j}(y)$ in Eq.~(\ref{delGS}) has the following form
\begin{equation}
\label{waight2} \Delta {\cal {W}}^{'}_{j}(y)=\Delta {\cal
{W}}^{}_{j}(y)\times \sum_ {m=0}^{5} {\cal A}_{m}y^{\frac
{m-1}{2}}\;.
\end{equation}\\
The additional term in the above equation, ($\sum$ term), serves
to control the behavior of the singlet sector at very low-$y$
values and the $\Delta {\cal {W}}^{}_{j}(y)$ defined in
Eq.~(\ref{waight1}).

 In the next sections we need to use the {\sc Mellin}
moments of Eqs.~(\ref{delGNS}, \ref{delGS}) to determine {\sc
Mellin} moments of PDFs in our analysis.

\section{The theoretical background of the QCD analysis}
Let us define the {\sc Mellin} moments for the polarized structure
function $g_1^p(x,Q^2)$:

\be g_1^p(N,Q^2)=\int_0^1 \;x^{N-1}g_1^p(x,Q^2)dx \label{momdefine}
\ee

 In the QCD-improved quark parton model (QPM),
i.e., at leading twist, and to leading logarithmic order in the
running strong coupling constant $\alpha_s(Q^2)$ of
Quantum-Chromodynamics (LO QCD), the deep-inelastic scattering off
the nucleon can be interpreted as the incoherent superposition of
virtual-photon interactions with quarks of any flavor $q$. By
angular momentum conservation, a spin-$\frac{1}{2}$
 parton can absorb a hard photon only when their
spin orientations are opposite.  The spin structure function $g_1$
has then a probabilistic interpretation, which for the proton
reads \cite{Altarelli:1977zs}
\begin{eqnarray}
g_1^p(N,Q^2)&=&\frac{1}{2} \sum_{q} e^2_q \left[ \Delta q(N,Q^2) +
\Delta \bar{q}(N,Q^2)\right]
\nonumber \\
&=&\frac{1}{2}\langle e^2\rangle \left[\Delta q_S (N,Q^2) +\Delta
q_{NS} (N,Q^2)\right]. \label{eq:momg1LO}
\end{eqnarray}
Here, the quantity $-Q^2$ is the squared four-momentum transferred
by the virtual photon, $N$ is the order of moments, $e_q$ is the
charge, in units of the elementary charge $\vert e \vert$, of quarks
of flavor $q$, $\langle e^2\rangle=\sum_q e^2_q/N_q$ is the average
squared charge of the $N_q$ active quark flavors, and
$\Delta{q}(N,Q^2)$ is the quark helicity distribution for
 quarks of flavor $q$. Correspondingly,
$\Delta{\bar{q}}(N,Q^2)$ is anti-quark helicity distributions. Moreover the
flavor singlet and flavor non-singlet quark helicity distributions
are defined as
\begin{eqnarray}\label{deltaq-s}
\Delta q_S(N,Q^2)&=&\sum_{q}\left[\Delta q(N,Q^2) + \Delta
\bar{q}(N,Q^2)\right] \nonumber \\
 &\equiv& \Delta {\Sigma}(N,Q^2),
\end{eqnarray}
and
\begin{eqnarray}\label{deltaq-ns}
\Delta q_{NS}(N,Q^2)\!\!=\!\!\frac{1}{\langle
e^2\rangle}\!\sum_qe^2_q\left[\Delta q(N,Q^2)\! +\!\Delta
\bar{q}(N,Q^2)\right]-\Delta q_S(N,Q^2).
\end{eqnarray}
For the analysis presented in this paper, only the three lightest
quark flavors, $q=u,d,s$, are taken into account  and the number of
active quark flavors $N_q$ is equal to three.\\

The twist--2 contributions to the structure function $g_1(N,Q^2)$
can be represented in terms of the polarized parton densities
 and the coefficient functions $\Delta C_i^N$ in the
{\sc Mellin} -N space  by \cite{Lampe:1998eu}

\begin{equation}
g_1^p(N,Q^2)=\frac{1}{2}\sum\limits_q
e^2_q\{(1+\frac{\alpha_s}{2\pi}\Delta C^N_q) [\Delta
q(N,Q^2)+\Delta\bar q(N,Q^2)] + \frac{\alpha_s}{2\pi}2\Delta
C^N_g\Delta g(N,Q^2)\}\;, \label{eq:momg1NLO}
\end{equation}
in this equation the NLO running coupling constant is given by
\begin{equation}
\alpha_{s}(Q^{2})\cong\frac{1}{b\log \frac{Q^{2}}{\Lambda_{\overline{MS}} ^{2}} }-\frac{%
b^{\prime }}{b^3}\frac{\ln \left( \ln  \frac{Q^{2}}{\Lambda_{\overline{MS}} ^{2}}%
 \right) }{\left( \ln \frac{Q^{2}}{\Lambda_{\overline{MS}} ^{2}} \right)
^{2}}\;,
\end{equation}
where $b=\frac{33-2f}{12\pi }$ and $b^{\prime }=\frac{153-19f}{24\pi
^{2}}$. In the above equations, we choose $Q_0=1\;GeV^2$ as a fixed
parameter and $\Lambda$ is an unknown parameter which can be
obtained  by fitting
to experimental data.\\

In Eq.~(\ref{eq:momg1NLO}), $\Delta
q(N,Q^2)=\Delta{{q_v(N,Q^2)}}+\Delta\bar q(N,Q^2)$, $\Delta\bar
q(N,Q^2)$ and $\Delta g (N,Q^2)$ are moments of the polarized parton
distributions in a proton. $\Delta C^N_q$, $\Delta C^N_g$ are also
the $N$-th moments of spin-dependent Wilson coefficients given by
\begin{equation}
\Delta C^N_q= \frac{4}{3}\biggl[-S_2(N)+(S_1(N))^2
+\left(\frac{3}{2}- \frac{1}{N(N+1)}\right) S_1(N) +\frac{1}{N^2}
+\frac{1}{2N}+\frac{1}{N+1}-\frac{9}{2} \biggr]\;,
\end{equation}
and
\begin{equation}
\Delta C^N_g=\frac{1}{2}[-\frac{N-1}{N(N+1)}(S_1(N)+1)
-\frac{1}{N^2}+ \frac{2}{N(N+1)}] \;, \end{equation}
with $S_{k}(N)$ defined as in Ref.~\cite{Lampe:1998eu}.\\

According to improved polarized valon model framework,
determination of the moments of parton distributions in a proton
can be done strictly through the moments of the polarized valon
distributions.

The moments of PPDFs are denoted respectively by:
$\Delta{u_{v}}(N,Q^{2})$, $\Delta{d_{v}}(N,Q^{2})$, $\Delta{\Sigma
}(N,Q^{2})$ and $\Delta g(N,Q^{2})$. Therefore, the moments of the
polarized $u$ and $d$-valence quark in a proton are convolutions of
two moments:
\begin{equation}
\Delta{u_{v}}(N,Q^{2})=2\Delta M^{NS}(N,Q^{2})\times\Delta
M^{'}_{U/p}(N)\;, \label{eq:momuv}
\end{equation}
\begin{equation}
\Delta{d_{v}}(N,Q^{2})=\Delta M^{NS}(N,Q^{2})\times\Delta M^{'}_
{D/p}(N)\;, \label{eq:momdv}
\end{equation}
In the above equation $M'_{j/p}(N)$ is the moment of $\Delta
G_{j/p}(y)$ distribution, i.e. $\Delta M'_{j/p}(N)=\int_{0}^{1}
y^{N-1}\;\Delta G_{j/p}(y)dy$.

The moment of the polarized singlet distribution ($\Sigma $) is as
follows:
\begin{equation}
\Delta{\Sigma }(N,Q^{2})=\Delta M^{S}(N,Q^{2})(2 \Delta
M^{''}_{U/p}(N)+\Delta M^{''}_{D/p}(N))\;. \label{eq:momsig}
\end{equation}
here $M''_{j/p}(N)$ is the moment of $\Delta G'_{j/p}(y)$
distribution, i.e. $\Delta M''_{j/p}(N)=\int_{0}^{1} y^{N-1}\;\Delta
G'_{j/p}(y)dy$. For the gluon distribution we have
\begin{equation}
\Delta g(N,Q^{2})=\Delta M^{gq}(N,Q^{2})(2 \Delta
M^{'}_{U/p}(N)+\Delta M^{'}_{D/p}(N))\;,
\end{equation}
where $\Delta M^{NS}(N,Q^{2})$ in
Eqs.~(\ref{eq:momuv},\ref{eq:momdv}), $\Delta M^{S}(N,Q^{2})$ in
Eq.~(\ref{eq:momsig}) and also $\Delta M^{gq}(N,Q^{2})$ which is
the quark-to-gluon evolution
function, are given in Ref. \cite{Khorramian:2004ih}\\

 It is obvious
that the final form for $g_1(N,Q^2)$ involves the total of 15
unknown parameters. If the parameters can be  obtained  then the
computation of all moments of the PPDFs and the polarized
structure
function (PSF), $g_1(N,Q^2)$, are possible. \\

\section{The method of the QCD analysis of PSF}
%\subsection{Jacobi polynomials approach}

 The evolution equations allow one to calculate the
$Q^2$-dependence of the PPD's provided  at a certain
reference point $Q_0^2$. These distributions are usually
parameterized on the basis of plausible theoretical assumptions
concerning their behavior near the end points $x=0,1$.

In the phenomenological investigations of the  polarized and
unpolarized structure functions, for example $xg_1^p$ or $xF_3^p$
for a given value of $Q^2$, only a limited number of experimental
points,
 covering a partial range of values of $x$, are available. Therefore,
one cannot directly determine the moments.
 A method devised to deal with this situation
 is to take  averages of the structure function weighted by suitable polynomials.

The evolution equation can be solved and QCD predictions for PSF
obtained with the help of various methods. For example we can use
the Bernstein polynomial to determine PPD's in the NLO
approximation to obtain some unknown parameters to parameterize
PPD's at $Q_0^2$. In this way, we can compare theoretical
predictions with the experimental results for the Bernstein
averages just in {\sc Mellin}-$N$ space. To obtain these
experimental averages from the E143 and SMC data
\cite{Abe:1998wq,Adams:1997tq}, we need to fit $x{g_1}(x,{Q^2})$
for each bin in ${Q}^{2}$ separately \cite{Khorramian:2004ih}.

 If we want to take into account very recent
experimental data \cite{unknown:2006vy} for all range value of
$Q^2$ to determine PPDFs, it is convenience to apply Jacobi
polynomials expansion and not the Bernstein one.
The advantage of application for all data points, especially very
recently HERMES experimental data \cite{unknown:2006vy},  and not
just a series of data points, is our motivation to use Jacobi
polynomials and not Bernstein polynomial to study spin dependent
of parton distribution function.

One of the simplest and fastest possibilities in the PSF
reconstruction from the QCD predictions for its Mellin moments is
Jacobi polynomials expansion. The Jacobi polynomials are
especially suited for this purpose since they allow one to factor
out an essential part of the $x$-dependence of the SF into the
weight function \cite{parisi}. Thus, given the Jacobi moments
$a_{n}(Q^2)$, a structure function $f(x,Q^2)$ may be reconstructed
in a form of the series \cite{Barker}-\cite{Barker:1980wu}
\begin{equation}
xf(x,Q^2)=x^{\beta}(1-x)^{\a} \sum_{n=0}^{N_{max}} a_{n}(Q^2)
\Theta_n ^{\a , \beta}(x), \label{e5}
\end{equation}
where $N_{max}$ is the number of polynomials and $\Theta_n ^{\a ,
\beta}(x)$ are the Jacobi polynomials of order $n$,
\begin{equation}
\Theta_{n} ^{\a , \beta}(x)= \sum_{j=0}^{n}c_{j}^{(n)}{(\a ,\beta
)}x^j , \label{e9}
\end{equation}
where $c_{j}^{(n)}{(\a ,\beta )}$ are the coefficients that
 expressed through $\Gamma$-functions and satisfy the orthogonality relation with the weight
$x^{\beta}(1-x)^{\a}$ as following

\begin{equation}
\int_{0}^{1}dx\;x^{\beta}(1-x)^{\a}\Theta_{k} ^{\a , \beta}(x)
\Theta_{l} ^{\a , \beta}(x)=\delta_{k,l}\ , \label{e8}
\end{equation}
For the moments, we note that the $Q^2$ dependence is entirely
contained in the Jacobi moments
\begin{eqnarray}
a_{n}(Q^2)&=&\int_{0}^{1}dx\;xf(x,Q^2)\Theta_{k} ^{\a ,
\beta}(x)\nonumber \\
&=&\sum_{j=0}^{n}c_{j}^{(n)}{(\a ,\beta )} f(j+2,Q^2) \;.
\label{e8nn}
\end{eqnarray}
obtained by inverting Eq.~\ref{e5}, using Eqs.~(\ref{e9},
\ref{e8}) and also definition of moments,
$f(j,Q^2)=\int_{0}^{1}dx\;x^{j-2}xf(x,Q^2)$.
\\
Using  Eqs. (\ref{e5}-\ref{e8nn}) now, one can relate the PSF with
its Mellin moments \ba
xg_{1}^{N_{max}}(x,Q^2)&=&x^{\beta}(1-x)^{\a}
\sum_{n=0}^{N_{max}}\Theta_n ^{\a, \beta}(x)
\sum_{j=0}^{n}c_{j}^{(n)}{(\a ,\beta )} g_{1}(j+2,Q^2),
\label{eg1Jacob} \ea where $g_{1}(j+2,Q^2)$ are the moments
determined by Eqs.(\ref{eq:momg1LO},\ref{eq:momg1NLO}). $N_{max}$,
$\alpha$ and $\beta$  have to be chosen so as to achieve the
fastest convergence of the series on the R.H.S. of
Eq.~(\ref{eg1Jacob}) and to reconstruct $xg_1$ with the required
accuracy. In our analysis we use $N_{max}=9$, $\alpha=3.0$ and
$\beta=0.5$. The same method has been applied to calculate the
nonsinglet structure function $xF_3$ from their moments [59-63].

 Obviously the $Q^2$-dependence
of the polarized structure function is defined by the
$Q^2$-dependence of the moments.

\section{The procedure of the QCD fit of  $g_1^p$ data}
The remarkable growth of experimental data on inclusive polarized
deep inelastic scattering of leptons off nucleons over the last
years allows to perform  refined QCD analyzes of polarized
structure functions in order to reveal the spin--dependent
partonic structure of the nucleon. For the QCD analysis presented
in the present paper the following data sets are used: the HERMES
proton data \cite{Airapetian:1998wi,unknown:2006vy}, the SMC
proton data \cite{Adeva:1998vv}, the E143 proton data
\cite{Abe:1998wq}, the EMC proton data
\cite{Ashman:1987hv,Ashman:1989ig}. The number of the published
data points above $Q^2 = 1.0~\GeV^2$ for the different data sets
are summarized in Table~1 for data on $g_1$ together with the $x-$
and $Q^2$--ranges for different experiments.
\\
\\
\begin{center}
\begin{tabular}{|c|c|c|c|c|}
\hline\hline Experiment & $x$--range &
\begin{tabular}{c}
$Q^{2}$--range \\
\lbrack GeV$^{2}$]
\end{tabular}
& number of data points & Ref. \\ \hline\hline E143(p) & 0.031 --
0.749  & 1.27 -- 9.52 & 28 & \cite{Abe:1998wq} \\
E143(p) & 0.031 -- 0.749 & 2, 3, 5 (Fixed)  & 84 & \cite{Abe:1998wq} \\
HERMES(p) &  0.028 -- 0.660   & 1.01 -- 7.36 & 19 & \cite{Airapetian:1998wi} \\
HERMES(p) & 0.023 -- 0.660 & 2.5 (Fixed) & 20 & \cite{Airapetian:1998wi} \\
HERMES(p) & 0.026 -- 0.731 & 1.12-14.29 & 62 & \cite{unknown:2006vy} \\
SMC(p) & 0.005 -- 0.480  & 1.30 -- 58.0 & 12 & \cite{Adeva:1998vv} \\

SMC(p) & 0.005 -- 0.480 & 10 (Fixed) & 12 & \cite{Adeva:1998vv} \\
EMC(p) & 0.015 -- 0.466  & 3.50 -- 29.5 & 10 &
\cite{Ashman:1987hv}
\\
EMC(p) & 0.015 -- 0.466 & 3.50 -- 29.5 & 10 & \cite{Ashman:1989ig}\\
\hline proton  &  &  & 257 &  \\ \hline\hline
\end{tabular}
\vspace{2mm} \noindent \\
{{\bf {Table~1:}} Published data points
above $Q^2 = 1.0$ GeV$^2$.}
\end{center}
 In the fitting
procedure we started with the 15 parameters selected, i.e. 4
parameters for each non-singlet polarized valon distribution, 6
parameters for singlet polarized valon distribution and
$\Lambda_{\rm QCD}$ to be determined. For this set of parameters
the sea--quark distribution was assumed to be described according
to $SU(3)$ flavor symmetry.

In the further procedure we fixed 2 parameters at their values
obtained in the first minimization and chose the first moment of
polarized Non-singlet valon distributions in the polarized
case~\cite{Khorramian:2004ih}. The lack of constraining power of
the present data on the polarized parton densities has to be
stressed, however. Since only more precise data can improve the
situation, so we add the recent experimental data achieved from HERMES
group\cite{unknown:2006vy}.

The final minimization was carried out under the above conditions
and determined the remaining 15 parameters. The values and errors
of these parameters along with those parameters fixed in the
parametrization, Sec.~2, are summarized in LO and NLO in table~2.
The results on $\Lambda_{\rm QCD}$ are discussed separately in
section~6. The starting scale of the evolution was chosen as
$Q^2_0 = 1$ GeV$^2$.

Using the CERN subroutine MINUIT \cite{James:1975dr}, we defined a
global ${\chi}^{2}$ for all the experimental data points  and
found an acceptable fit with minimum
${\chi}^{2}/{\rm{d.o.f.}}=236.577/242=0.978$ in the LO case and
${\chi}^{2}/{\rm{d.o.f.}}=225.920/242=0.933$ in the NLO case. In this table we compare
the results reported in \cite{Khorramian:2004ih} which are based on Bernstein approach
and the results of the present analysis. Also we obtain the uncertainties of
the parameters in
 Jacobi Approach which are not calculated in Bernstein approach.
 We should notice that the Jacobi and Bernstein polynomials are
merely used as a tool in the fitting procedure and the results are
independent of it.

\begin{center}
\begin{tabular}{|c|c|c|c|c|}
\hline\hline & \multicolumn{2}{|c|}{Bernstein Approach \cite{Khorramian:2004ih}} &
\multicolumn{2}{|c|}{Jacobi Approach} \\ \cline{2-5}& LO & NLO & LO &
NLO \\ \cline{2-5} & value & value & value & value \\ \hline\hline
$\Lambda _{QCD}^{(4)};$MeV & 203 & 235 & 201$\pm$110 & 245$\pm$58 \\ \hline\hline
$N_{_U}(=\xi_{_U} A_{_U})$ & 0.0020 & 0.0038 & 0.0015(fixed) & 0.0018(fixed) \\
$\alpha _{_U}$ & -2.3789 & -2.1501 & -2.4489$\pm$0.024 & -2.3590$\pm$0.028 \\
$\beta _{_U}$ & -1.7518 & -0.8859 & -1.9050$\pm$0.196 & -0.9683$\pm$ 0.206\\
$\gamma _{_U}$ & 11.0804 & 10.6537 & 13.5420$\pm$0.378 & 22.6095$\pm$ 0.418\\
$\eta _{_U}$ & -1.4629 & -0.1548 & -1.8339$\pm$0.120& -1.9778$\pm$ 0.150\\
\hline\hline
$N_{_D}(=\xi_{_{_D}} A_{_{_D}})$ & -0.005 & -0.0046 & -0.0029(fixed) & -0.0026(fixed) \\
$\alpha _{_D}$ & -1.5465 & -1.5859 & -1.6059$\pm$0.029 & -1.6638$\pm$0.034 \\
$\beta _{_D}$ & -1.8776 & -1.5835 & -2.2009$\pm$0.353 & -1.6214$\pm$ 0.453\\
$\gamma _{_D}$ & 8.5042 & 9.6205 & 10.2371$\pm$0.829 & 13.1966$\pm$ 0.942\\
$\eta _{_D}$ & -0.8608 & -0.8410 & 0.8751$\pm$0.091 & 0.8483$\pm$0.142 \\
\hline\hline
$\mathcal{A}_{0}$ & 0.0004 & -0.0025 & 0.0003$\pm$ 10$^{-5}$ & 0.0052$\pm$10$^{-5}$ \\
$\mathcal{A}_{1}$ & 0.2954 & -3.1148 & 0.2611$\pm$0.191 & -3.0034$\pm$0.241 \\
$\mathcal{A}_{2}$ & -6.9134 & 15.8114 & -6.5429$\pm$0.174 & 14.8696$\pm$ 0.214\\
$\mathcal{A}_{3}$ & 30.9851 & -21.1500 & 29.6999$\pm$0.475 & -18.8057$\pm$ 0.051\\
$\mathcal{A}_{4}$ & -39.7383 & 10.5025 & -37.9379$\pm$0.202 & 8.4598$\pm$ 0.298\\
$\mathcal{A}_{5}$ & 16.4605 & -0.9162 & 15.5879$\pm$0.041& -0.3410$\pm$0.050 \\
\hline\hline $\chi ^{2}/ndf$ &
\multicolumn{1}{|c}{154.98/123=1.260} & \multicolumn{1}{|c}{
115.62/123=0.940} & \multicolumn{1}{|c}{236.577/242=0.978}
 & \multicolumn{1}{|c|}{225.920/242=0.933} \\
\hline\hline
\end{tabular}\\
\end{center}
{{{\bf {Table~2:}} Parameter values in LO and NLO of the parameter
fit for Bernstein and Jacobi approaches.}}

\section{Results and conclusions of the QCD analysis}

We have performed a QCD analysis of the inclusive polarized
deep--inelastic charged lepton--nucleon scattering  data to
next--to--leading order and derived parameterizations of polarized
valon distributions at a starting scale $Q_0^2$ together with the
QCD--scale $\Lambda_{\rm QCD}$ in the polarized valon model
framework.

The  analysis was performed using the Jacobi polynomials--method
to determine the parameters of the problem in a fit to the data.

A new aspect in comparison with previous analyzes is that we
determine the parton densities and the QCD scale in leading and
next--to--leading order by using Jacobi polynomial expansion
method.

 Detailed
comparisons were performed to the results obtained in other recent
parameterizations~\cite{Goto:1999by, Blumlein:2002be,
Gluck:2000dy, Leader:2005ci, deFlorian:2005mw}. The previous
results are widely compatible with the present parameterizations.
These distributions can be used in the numerical calculations for
polarized high--energy scattering processes at hadron-- and
$ep$--colliders.

Looking at the $Q^2$ dependence of the structure function
$g_1(x,Q^2)$ in intervals of $x$ gives insight to the scaling
violations in the spin sector. As in the unpolarized case the
presence of scaling violations are expected to manifest in a slope
changing with $x$. The proton data on $g_1(x,Q^2)$ have been
plotted in such a way in Figure~1 and confronted with the QCD NLO
curves of the present analysis. Corresponding curves of the
parameterizations~\cite{Goto:1999by, Blumlein:2002be,
Gluck:2000dy, Leader:2005ci, deFlorian:2005mw} are also shown.
Slight but non-significant differences between the different
analyzes are observed in the intervals at low values of $x$.
However, the data are well covered within the errors by all
analyzes.\\

\begin{figure}[tbh]
\centerline{\includegraphics[width=0.45\textwidth]{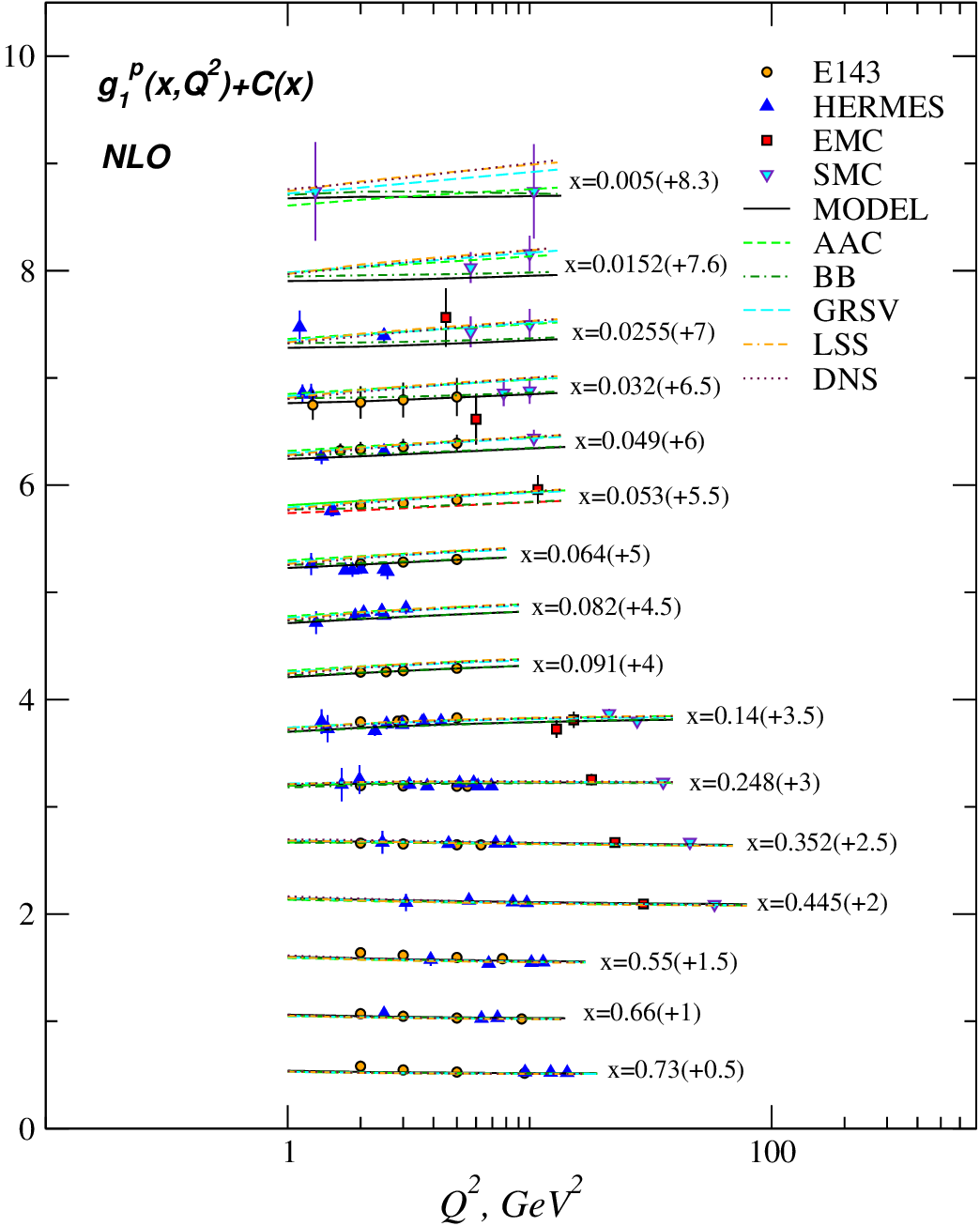}}
\caption{The polarized structure function $g_1^p$ as function of
$Q^2$ in intervals of $x$. The error bars shown are the
statistical and systematic uncertainties added in quadrature. The
data are well described by our QCD NLO curves (solid lines) which
based on the valon model and Jacobi polynomials expansion method.
The values of $C(x)$ are given in parentheses. Also shown are the
QCD NLO curves obtained by AAC (dashed lines) \cite{Goto:1999by},
BB (dashed-dotted lines) \cite{Blumlein:2002be}, GRSV (long-dashed
lines) \cite{Gluck:2000dy}, LSS (dashed-dashed-dotted lines)
\cite{Leader:2005ci} and DNS (dotted lines)
\cite{deFlorian:2005mw} for comparison.}
\end{figure}
In Figures~2--3 the fitted parton distribution functions,  in
leading and next-to-leading order for all sets of
parameterizations ~\cite{Goto:1999by, Blumlein:2002be,
Gluck:2000dy, Leader:2005ci} and their errors are presented at the
starting scale $Q_0^2$.

\begin{figure}[tbh]
\centerline{\includegraphics[width=0.6\textwidth]{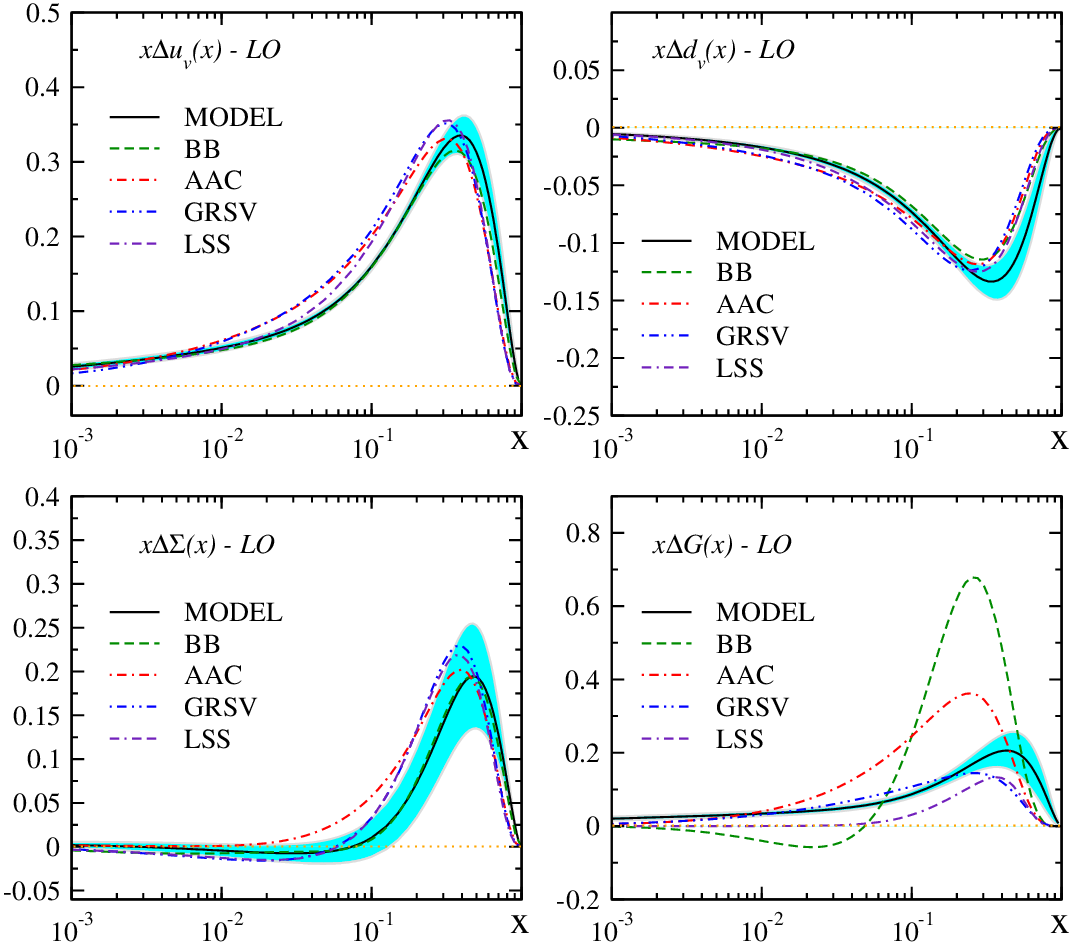}}
\caption{ LO polarized parton distributions at the input scale
$Q_0^2$=1.0 GeV$^2$ compared to results obtained by BB model
(dashed line) (ISET=1)\cite{Blumlein:2002be}, AAC (dashed-dotted
line) (ISET=1)\cite{Goto:1999by}, GRSV (dashed-dotted dotted line)
(ISET=3)\cite{Gluck:2000dy} and LSS (dashed-dashed dotted line)
(ISET=3)\cite{Leader:2005ci} }
\end{figure}
\begin{figure}[tbh]
\vspace{1 cm}
\centerline{\includegraphics[width=0.6\textwidth]{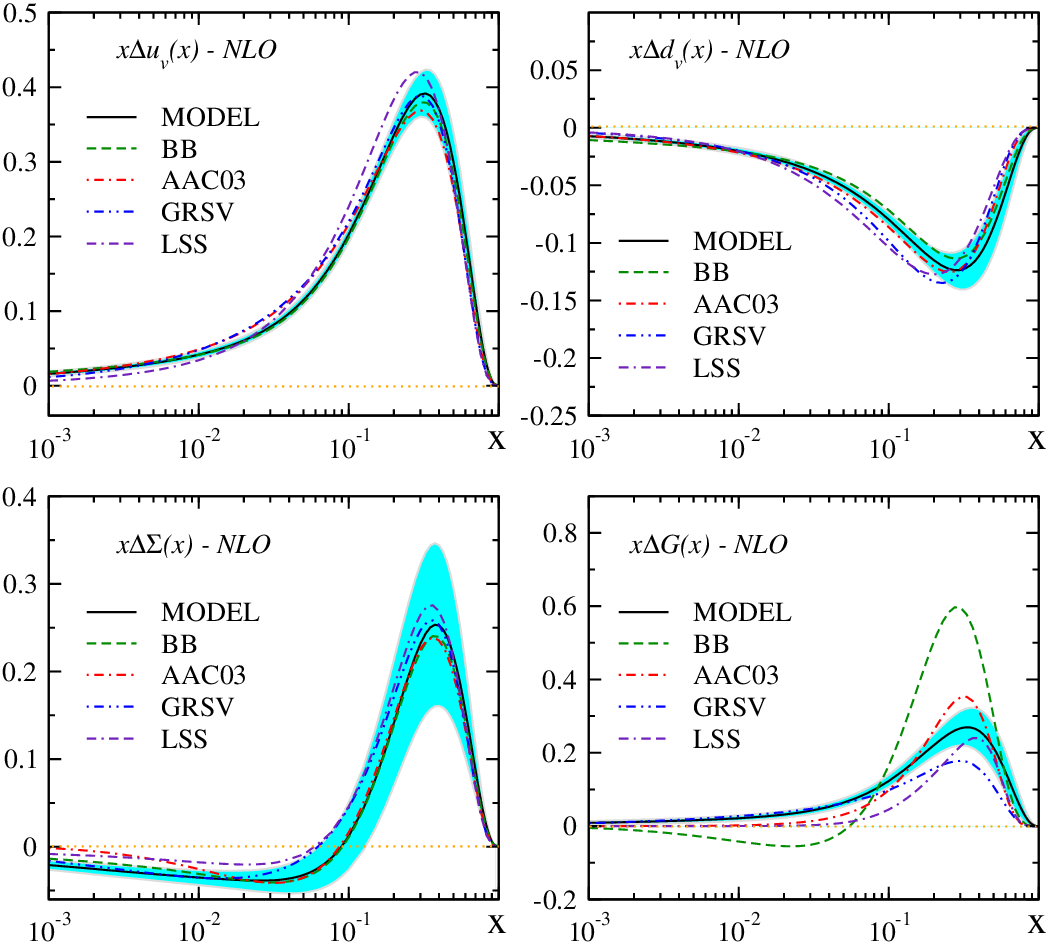}}
\caption{ NLO polarized parton distributions at the input scale
$Q_0^2$=1.0 GeV$^2$ compared to results obtained by BB model
(dashed line) (ISET=3)\cite{Blumlein:2002be}, AAC (dashed-dotted
line) (ISET=3)\cite{Goto:1999by}, GRSV (dashed-dotted dotted line)
(ISET=1)\cite{Gluck:2000dy} and LSS (dashed-dashed dotted line)
(ISET=1)\cite{Leader:2005ci}}
\end{figure}

The polarized structure function $xg_1^p(x.Q^2)$
measured in the interval $3.0$ GeV$^2 < Q^2 < 5.0$ GeV$^2$,
Figure~4, using the world asymmetry data is well described by our
QCD NLO curve. We also compare to corresponding representations of
the parameterizations \cite{Leader:2005ci, Gluck:2000dy,
Blumlein:2002be, Goto:1999by}, which are compatible within the
present results.

In Figures~5--8 the scaling violations of the individual polarized
momentum densities are depicted in the range
$x~\epsilon~[10^{-3},1],~~ Q^2~\epsilon~[1,10^4]~{\,\mbox{GeV}}$
choosing the NLO distributions. The up-valence distribution
$x\Delta u_v$, Figure~5, evolves towards smaller values of $x$ and
the peak around $x \sim 0.25$ becomes more flat in the evolution
from $Q^2=1 \GeV^2$ to $Q^2 = 10^4 \GeV^2$. Statistically this
distribution is constrained best among all others. The
down-valence distribution $x \Delta d_v$, Figure 6, remains
negative in the same range, although it is less constraint by the
present data than the up-valence density. Also here the evolution
is towards smaller values of $x$ and structures at larger $x$
flatten out.

\begin{figure}[tbh]
\vspace{1 cm}
\centerline{\includegraphics[width=0.5\textwidth]{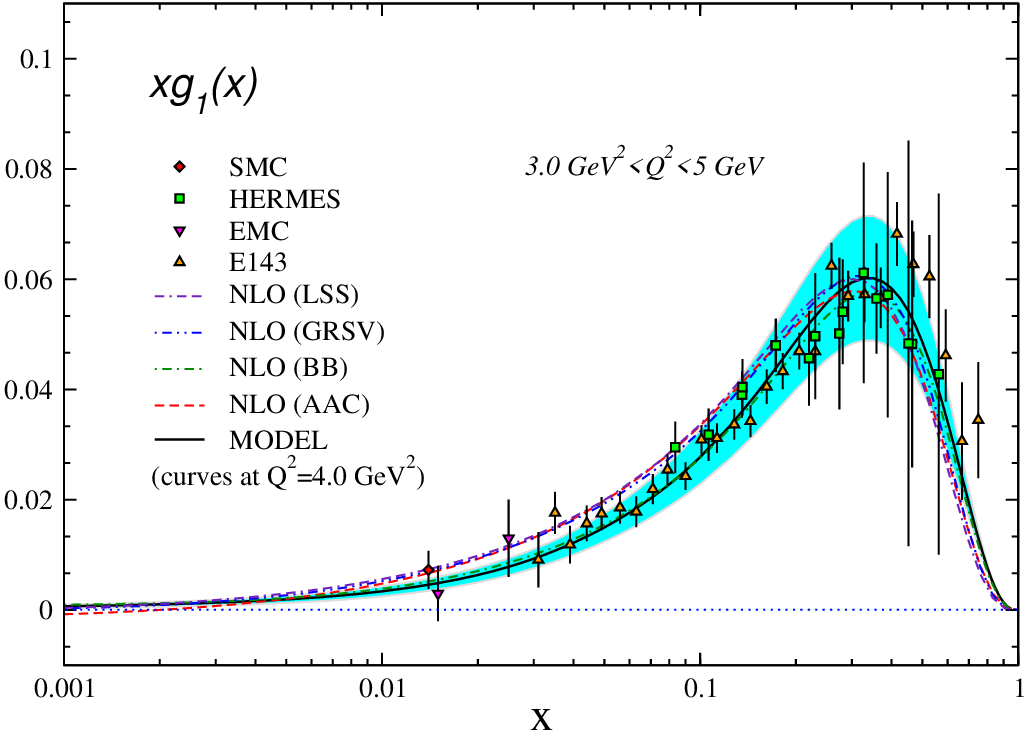}}
\caption{Polarized proton structure function $xg_{1}^{p}$ measured
in the interval 3.0 GeV$^2< Q^2<$ 5.0 GeV$^2$ as a function of
$x$. Also shown are the QCD NLO curves at the same value of Q$^2$
obtained by LSS (dashed-dashed dotted line)
(ISET=1)\cite{Leader:2005ci}, GRSV (dashed-dotted dotted line)
(ISET=1)\cite{Gluck:2000dy}, BB model (dashed-dotted line)
(ISET=3)\cite{Blumlein:2002be} and AAC (dashed line)
(ISET=3)\cite{Goto:1999by} for comparison.}
\end{figure}
\begin{figure}[tbh]
\vspace{1 cm}
\centerline{\includegraphics[width=0.5\textwidth]{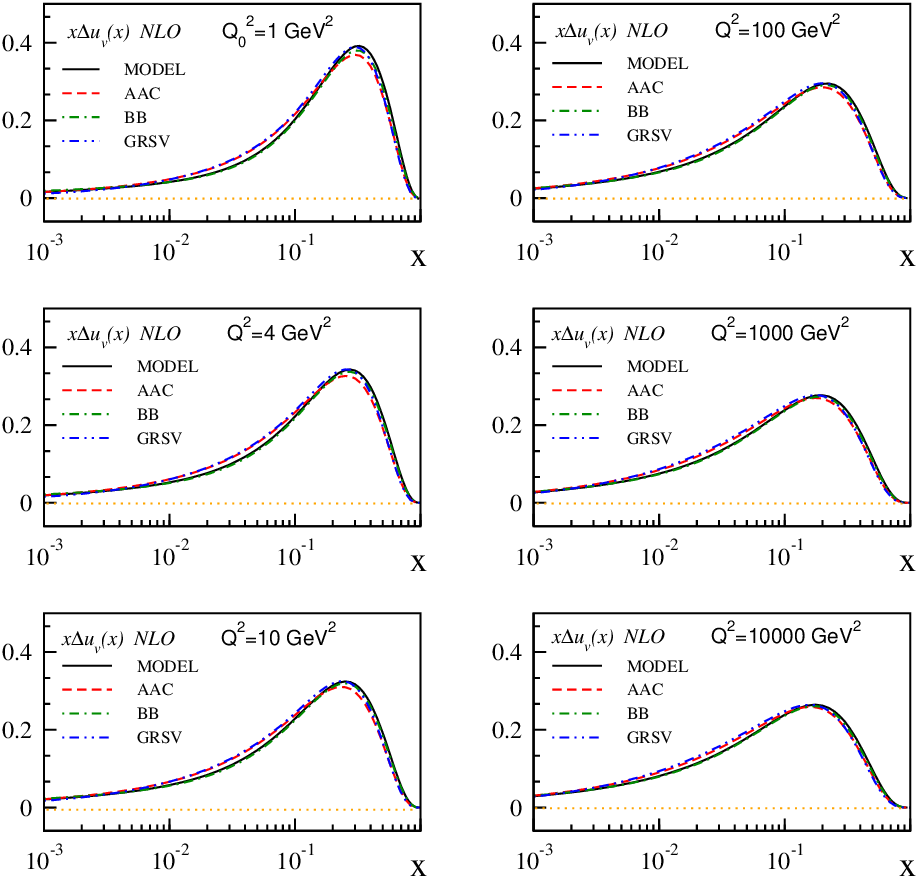}}
\caption{The Polarized parton distribution $x \Delta u_v$, evolved
up to values of $Q^2$=10,000 GeV$^2$ as a function of $x$ in the
NLO approximation. The solid line is our model, dashed line is the
AAC model (ISET=3)\cite{Goto:1999by}, dashed-dotted line is the BB
model (ISET=3)\cite{Blumlein:2002be} and dashed-dotted-dotted line
is the GRSV model (ISET=1)\cite{Gluck:2000dy}.}
\end{figure}
\begin{figure}[tbh]
\vspace{1 cm}
\centerline{\includegraphics[width=0.5\textwidth]{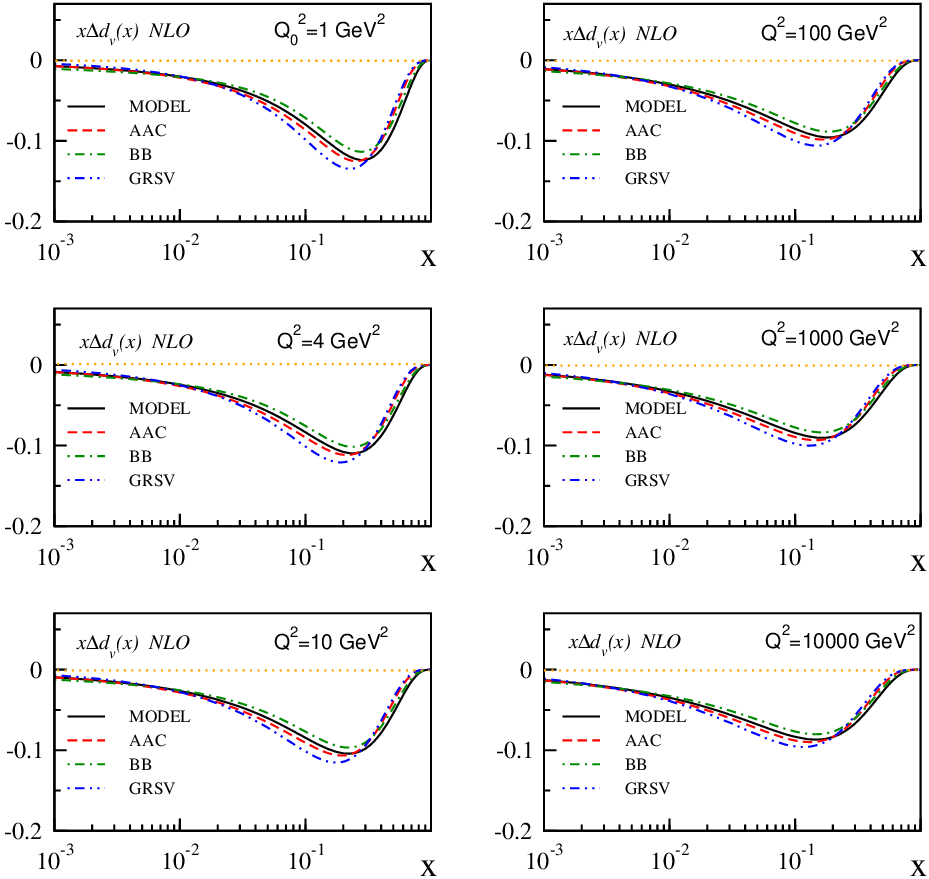}}
\caption{The Polarized parton distribution $x \Delta d_v$, evolved
up to values of $Q^2$=10,000 GeV$^2$ as a function of $x$ in the
NLO approximation. The solid line is our model, dashed line is the
AAC model (ISET=3)\cite{Goto:1999by}, dashed-dotted line is the BB
model (ISET=3)\cite{Blumlein:2002be} and dashed-dotted-dotted line
is the GRSV model (ISET=1)\cite{Gluck:2000dy}.}
\end{figure}
The momentum density of the polarized singlet
quark $x \Delta \Sigma$, Figure~7, is positive in the kinematic
range shown for all $Q^2$ and for $x \geq 0.1$ ,  but changes sign
for lower values of $x$. The maximum of the distribution at $Q^2 =
1 \GeV^2$ around $x \sim 0.3$ moves to $x \sim 0.2$ at $Q^2 = 10^4
{\,\mbox{GeV}}$. At the same time a minimum around $x \sim 0.03$
moves to $x \sim 0.002$.

The momentum density of the polarized gluon $x \Delta G$,
Figure~8,
 is positive in the kinematic range shown for all $Q^2$.
 Also in  this  case the evolution moves the
shape towards lower values of $x$ and flattens the distribution.
\begin{figure}[tbh]
\vspace{0.25 cm}
\centerline{\includegraphics[width=0.5\textwidth]{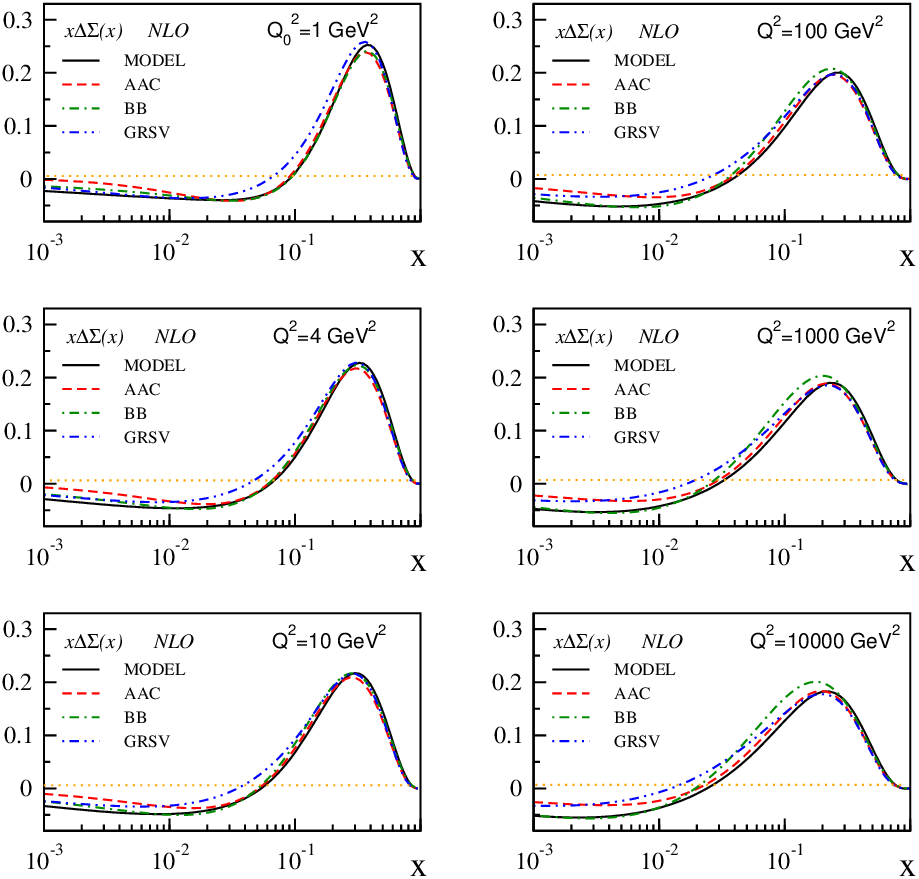}}
\caption{The Polarized parton distribution $x \Delta \Sigma$,
evolved up to values of $Q^2$=10,000 GeV$^2$ as a function of $x$
in the NLO approximation. The solid line is our model, dashed line
is the AAC model (ISET=3)\cite{Goto:1999by}, dashed-dotted line is
the BB model (ISET=3)\cite{Blumlein:2002be} and
dashed-dotted-dotted line is the GRSV model
(ISET=1)\cite{Gluck:2000dy}.}
\end{figure}
\begin{figure}[tbh]
\centerline{\includegraphics[width=0.5\textwidth]{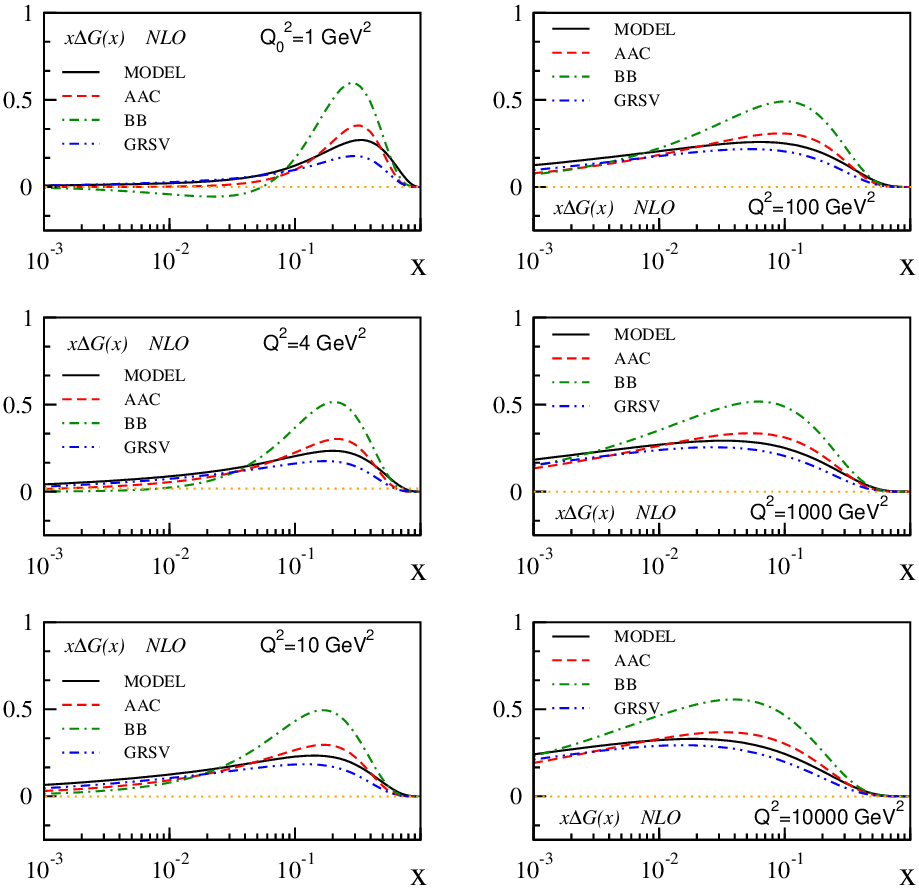}}
\caption{The Polarized parton distribution $x \Delta G$, evolved
up to values of $Q^2$=10,000 GeV$^2$ as a function of $x$ in the
NLO approximation. The solid line is our model, dashed line is the
AAC model (ISET=3)\cite{Goto:1999by}, dashed-dotted line is the BB
model (ISET=3)\cite{Blumlein:2002be} and dashed-dotted-dotted line
is the GRSV model (ISET=1)\cite{Gluck:2000dy}.}
\end{figure}
 By having polarized parton
distributions, the first moments of the polarized parton
distributions can be obtain. The first moments of the polarized
parton densities in NLO in the $\overline{\rm MS}$ scheme at $Q^2=
4\GeV^2$ for different sets of recent parton parameterizations are
presented in table~3.
\renewcommand{\arraystretch}{1.3}
\begin{center}
\begin{tabular}{|c|c|c|c|c|c|c|c|}
\hline \hline
%           &      &       &                &       &                \\
\multicolumn{1}{|c|}{Distribution}& \multicolumn{1}{ c|}{{
Model}}& \multicolumn{1}{ c|}{{Ref.~\cite{Khorramian:2004ih}}}&
\multicolumn{1}{ c|}{BB~\cite{Blumlein:2002be}}& \multicolumn{1}{
c|}{GRSV~\cite{Gluck:2000dy}}&
\multicolumn{1}{ c|}{AAC~\cite{Goto:1999by}}\\
\hline\hline $\Delta u_v$  & $0.9207 $
              & $0.8769$
&    0.926            & 0.9206      & 0.9278      \\
$\Delta d_v$  & --0.3391 &--0.3313 &--0.341 & --0.3409      &
--0.3416
\\
$\Delta u                      $ & $0.8390$ & $0.8145$
& 0.851  &  0.8593      &   0.8399   \\
$\Delta d                      $ & --0.4207 & --0.3937
& --0.415               & --0.4043      & --0.4295   \\
$\Delta \overline{q}$ &       --0.0817 &   --0.0624 & -0.074
&--0.0625 &--0.0879    \\
$\Delta G  $  & $0.9325$ &       $0.8322 $
&     $1.026$    & 0.6828    &  0.8076      \\
\hline\hline
\end{tabular}
\end{center}
\normalsize \vspace{2 mm} \noindent {{\bf {Table~3:}} Comparison
of the first moments of the polarized parton densities in NLO in
the $\overline{\rm MS}$ scheme at $Q^2= 4\GeV^2$ for different
sets of recent parton parameterizations. The second column (Model)
contains the first moments which is obtained from the valon model
and the Jacobi polynomials expansion method. }
\renewcommand{\arraystretch}{1}
\\

We can also obtain the first moment of $g_{1}^p$ (the Ellis-Jaffe
sum rule) by \be \Gamma _{1}^p(Q^{2})\equiv
\int_{0}^{1}dxg_{1}^p(x,Q^{2})\;. \ee  The results have also been
given in table~4. The first moments of polarized parton
distributions also shown for some value of $Q^{2}$.
\begin{center}
\begin{tabular}{ccccccc}
\hline\hline
$Q^{2}(GeV^{2})$ & $\Delta u_{v}$ & $\Delta d_{v}$ & $\Delta \Sigma $ & $%
\Delta \overline{q}$ & $\Delta g$ & $\Gamma_1^p$ \\ \hline
1  & 0.9260 & -0.3410 & 0.0965 & -0.0814 & 0.5850 & 0.1161 \\
3  & 0.9215 & -0.3393 & 0.0923 & -0.0816 & 0.8651 & 0.1195 \\
5 & 0.9202 & -0.3389 & 0.0911 & -0.0817 & 0.9837 & 0.1205 \\
10 & 0.9189 & -0.3384 & 0.0898 & -0.0818 & 1.1383 & 0.1215 \\
\hline\hline
\end{tabular}
\end{center}
{{{\bf {Table~4:}} The first moments of polarized parton
distributions, $\Delta u_{v}$, $\Delta d_{v}$, $\Delta \Sigma $,
$\Delta \overline{q}$, $\Delta g$ and $\Gamma_1^p$  in the  NLO
approximation for some value of $Q^{2}$.}}
\\
\\
In the framework
of QCD the spin of the proton can be expressed in terms of the
first moment of the total quark and gluon helicity distributions
and their orbital angular momentum, i.e.
 \be
  \frac{1}{2}=\frac{1}{2}\Delta \Sigma
^{p}+\Delta g^{p}+L_{z}^{p}, \ee where $L_{z}^{p}$ is the total orbital
angular momentum of all the quarks and gluons.\\
The contribution of addition of $\frac{1}{2}\Delta \Sigma$ and $\Delta g$
for typical value of
 $Q^2=4$ GeV$^2$ is around 0.978 in our analysis.
We can also compare this value in NLO with other recent analysis.
The reported value from BB model \cite{Blumlein:2002be} is 1.096,
AAC model \cite{Goto:1999by} is 0.837 and
 also GRSV model \cite{Gluck:2000dy} is 0.785. \\

In the QCD analysis we parameterized the strong coupling constant
$\alpha_s$ in terms of four massless flavors determining
$\Lambda_{\rm QCD}$. The LO and NLO results fitting the data, are
\begin{eqnarray}
\Lambda_{\rm QCD}^{(4)\rm \overline{MS}} &=& 201 \pm 110 (stat)
\; \MeV,~~{\tt LO}, \nonumber \\
\Lambda_{\rm QCD}^{(4)\rm \overline{MS}} &=& 245 \pm 58 (stat) \;
\MeV,~~{\tt NLO},
\end{eqnarray}
These results can be expressed in terms of $\alpha_s(M_Z^2)$:
\begin{eqnarray}
\alpha_s(M_Z^2) = 0.1281 \pm 0.0094 (stat)
\; \MeV,~~{\tt LO},  \nonumber \\
\alpha_s(M_Z^2) = 0.1141 \pm 0.0036 (stat)\; \MeV,~~{\tt NLO}.
\end{eqnarray}
These values can be compared with results from other QCD analyzes of polarized
inclusive deep--inelastic scattering data

\vspace*{-0.25cm}
%-------------------------------------------------------------------------
\begin{eqnarray}
{\rm E154}~\cite{abe97}: \quad \alpha_s(M_Z^2) & = & 0.108-0.116~,  \nonumber \\
{\rm SMC}~\cite{Adeva:1998vw}: \quad \alpha_s(M_Z^2) & = & 0.121 \pm 0.002 (stat)~,  \nonumber \\
{\rm ABFR}~\cite{alt98abfr}: \quad \alpha_s(M_Z^2) & = &
0.120\begin{array}{c} + 0.004 \\ - 0.005
\end{array} {\rm (exp)}
\begin{array}{c} + 0.009 \\ - 0.006 \end{array} {\rm (theor)}~, \nonumber \\
{\rm BB}~\cite{Blumlein:2002be}: \quad \alpha_s(M_Z^2) & = &
0.113\pm 0.004 (stat),\;\;\;\hbox{(ISET=3)~,}
\end{eqnarray}
%-------------------------------------------------------------------------
\noindent
and with the value of the current world average
%-------------------------------------------------------------------------
\begin{eqnarray}
 \alpha_s(M_Z^2) = 0.118 \pm 0.002~~\cite{Groom:2000in}~.
\end{eqnarray}
%-----------------
 We hope our results of QCD analysis of structure functions in terms of Jacobi polynomials
 could be able to describe more complicated hadron structure functions. We also
hope to be able to consider the symmetry breaking of polarized sea
quarks by using the polarized structure function expansion in the Jacobi polynomials.

%\newpage
\section{Acknowledgments}

We are especially grateful to G. Altarelli for fruitful
suggestions and critical remarks. A.N.K. is grateful to A. L.
Kataev for useful discussions and remarks during the visit to
ICTP. I am grateful to the staff of this center for providing
excellent conditions for work. A.N.K is grateful to CERN for their
hospitality whilst he visited there and could amend this paper. We
would like to thank M. Ghominejad and Z. Karamloo  for reading the
manuscript of this paper. We acknowledge the Institute for Studies
in Theoretical Physics and Mathematics (IPM) for financially
supporting this project. S.A.T. thanks Persian Gulf university for
partial financial support of this project.

\end{document}